
%
\input amstex
%
%
%
%
%
%
%


\magnification=1200
\hsize=31pc
\vsize=55 truepc
\hfuzz=2pt
\vfuzz=4pt
\pretolerance=500
\tolerance=500
\parskip=0pt plus 1pt
\parindent=16pt
%

%
%
\font\fourteenrm=cmr10 scaled \magstep2
\font\fourteeni=cmmi10 scaled \magstep2
\font\fourteenbf=cmbx10 scaled \magstep2
\font\fourteenit=cmti10 scaled \magstep2
\font\fourteensy=cmsy10 scaled \magstep2

%
\font\large=cmbx10 scaled \magstep1

%

%

%

%
\font\eightrm=cmr8
\font\eighti=cmmi8
\font\eightbf=cmbx8
\font\eightit=cmti8

\font\eightsy=cmsy8
\font\sixrm=cmr6
\font\sixi=cmmi6
\font\sixsy=cmsy6

%
\def\tenpoint{\def\rm{\fam0\tenrm}%
  \textfont0=\tenrm \scriptfont0=\sevenrm
                      \scriptscriptfont0=\fiverm
  \textfont1=\teni  \scriptfont1=\seveni
                      \scriptscriptfont1=\fivei
  \textfont2=\tensy \scriptfont2=\sevensy
                      \scriptscriptfont2=\fivesy
  \textfont3=\tenex   \scriptfont3=\tenex
                      \scriptscriptfont3=\tenex
  \textfont\itfam=\tenit  \def\it{\fam\itfam\tenit}%
  \textfont\slfam=\tensl  \def\sl{\fam\slfam\tensl}%
  \textfont\bffam=\tenbf  \scriptfont\bffam=\sevenbf
                            \scriptscriptfont\bffam=\fivebf
                            \def\bf{\fam\bffam\tenbf}%
  \normalbaselineskip=20 truept
  \setbox\strutbox=\hbox{\vrule height14pt depth6pt
width0pt}%
  \let\sc=\eightrm \normalbaselines\rm}
\def\eightpoint{\def\rm{\fam0\eightrm}%
  \textfont0=\eightrm \scriptfont0=\sixrm
                      \scriptscriptfont0=\fiverm
  \textfont1=\eighti  \scriptfont1=\sixi
                      \scriptscriptfont1=\fivei
  \textfont2=\eightsy \scriptfont2=\sixsy
                      \scriptscriptfont2=\fivesy
  \textfont3=\tenex   \scriptfont3=\tenex
                      \scriptscriptfont3=\tenex
  \textfont\itfam=\eightit  \def\it{\fam\itfam\eightit}%
  \textfont\bffam=\eightbf  \def\bf{\fam\bffam\eightbf}%
  \normalbaselineskip=16 truept
  \setbox\strutbox=\hbox{\vrule height11pt depth5pt width0pt}}
\def\fourteenpoint{\def\rm{\fam0\fourteenrm}%
  \textfont0=\fourteenrm \scriptfont0=\tenrm
                      \scriptscriptfont0=\eightrm
  \textfont1=\fourteeni  \scriptfont1=\teni
                      \scriptscriptfont1=\eighti
  \textfont2=\fourteensy \scriptfont2=\tensy
                      \scriptscriptfont2=\eightsy
  \textfont3=\tenex   \scriptfont3=\tenex
                      \scriptscriptfont3=\tenex
  \textfont\itfam=\fourteenit  \def\it{\fam\itfam\fourteenit}%
  \textfont\bffam=\fourteenbf  \scriptfont\bffam=\tenbf
                             \scriptscriptfont\bffam=\eightbf
                             \def\bf{\fam\bffam\fourteenbf}%
  \normalbaselineskip=24 truept
  \setbox\strutbox=\hbox{\vrule height17pt depth7pt width0pt}%
  \let\sc=\tenrm \normalbaselines\rm}
\def\today{\number\day\ \ifcase\month\or
  January\or February\or March\or April\or May\or June\or
  July\or August\or September\or October\or November\or
December\fi
  \space \number\year}
\def\monthyear{\ifcase\month\or
  January\or February\or March\or April\or May\or June\or
  July\or August\or September\or October\or November\or
December\fi
  \space \number\year}

%
\newcount\secno      
\newcount\subno      
\newcount\subsubno   
\newcount\appno      
\newcount\tableno    
\newcount\figureno   
%

%
\normalbaselineskip=20 truept
\baselineskip=20 truept

%
%
\def\title#1
   {\vglue1truein
   {\baselineskip=24 truept
    \pretolerance=10000
    \raggedright
    \noindent \fourteenpoint\bf #1\par}
    \vskip1truein minus36pt}
%

%
\def\author#1
  {{\pretolerance=10000
    \raggedright
    \noindent {\large #1}\par}}

%
\def\address#1
   {\bigskip
    \noindent \rm #1\par}

%
\def\shorttitle#1
   {\vfill
    \noindent \rm Short title: {\sl #1}\par
    \medskip}

%
\def\pacs#1
   {\noindent \rm PACS number(s): #1\par
    \medskip}

%
\def\jnl#1
   {\noindent \rm Submitted to: {\sl #1}\par
    \medskip}

%
\def\date
   {\noindent Date: \today\par
    \medskip}

%

%
\def\keyword#1
   {\bigskip
    \noindent {\bf Keyword abstract: }\rm#1}

%

%
%

%
\def\entry#1#2#3
   {\noindent
    \hangindent=20pt
    \hangafter=1
    \hbox to20pt{#1 \hss}#2\hfill #3\par}

%
\def\subentry#1#2#3
   {\noindent
    \hangindent=40pt
    \hangafter=1
    \hskip20pt\hbox to20pt{#1 \hss}#2\hfill #3\par}
\def\checkforsub{\futurelet\nexttok\decide}
\def\ssf{\relax}
\def\decide{\if\nexttok\ssf\let\endspace=\nospace
                \else\let\endspace=\extraspace\fi\endspace}
\def\nospace{\nobreak\par\nobreak}
%
%
\def\section#1{%
    \goodbreak
    \vskip24pt plus12pt minus12pt
    \nobreak
    \gdef\extraspace{\nobreak\bigskip\noindent\ignorespaces}%
    \noindent
    \subno=0 \subsubno=0
    \global\advance\secno by 1
    \noindent {\bf \the\secno. #1}\par\checkforsub}

%
\def\subsection#1{%
     \goodbreak
     \vskip24pt plus12pt minus6pt
     \nobreak
     \gdef\extraspace{\nobreak\medskip\noindent\ignorespaces}%
     \noindent
     \subsubno=0
     \global\advance\subno by 1
     \noindent {\sl \the\secno.\the\subno. #1\par}\checkforsub}

%
\def\subsubsection#1{%
     \goodbreak
     \vskip15pt plus6pt minus6pt
     \nobreak\noindent
     \global\advance\subsubno by 1
     \noindent {\sl \the\secno.\the\subno.\the\subsubno. #1}\null.
     \ignorespaces}

%
\def\appendix#1
   {\vskip0pt plus.1\vsize\penalty-250
    \vskip0pt plus-.1\vsize\vskip24pt plus12pt minus6pt
    \subno=0
    \global\advance\appno by 1
    \noindent {\bf Appendix \the\appno. #1\par}
    \bigskip
    \noindent}

%
\def\subappendix#1
   {\vskip-\lastskip
    \vskip36pt plus12pt minus12pt
    \bigbreak
    \global\advance\subno by 1
    \noindent {\sl \the\appno.\the\subno. #1\par}
    \nobreak
    \medskip
    \noindent}

%


%

%
\def\tabcaption#1
   {\global\advance\tableno by 1
    \noindent {\bf Table \the\tableno.} \rm#1\par
    \bigskip}

%

%

%

%

%

%
\def\figcaption#1
   {\global\advance\figureno by 1
    \noindent {\bf Figure \the\figureno.} \rm#1\par
    \bigskip}

%

%

%
\def\refjl#1#2#3#4
   {\hangindent=16pt
    \hangafter=1
    \rm #1
   {\frenchspacing\sl #2
    \bf #3}
    #4\par}

%
\def\refbk#1#2#3
   {\hangindent=16pt
    \hangafter=1
    \rm #1
   {\frenchspacing\sl #2}
    #3\par}

%
\def\numrefjl#1#2#3#4#5
   {\parindent=40pt
    \hang
    \noindent
    \rm {\hbox to 30truept{\hss #1\quad}}#2
   {\frenchspacing\sl #3\/
    \bf #4}
    #5\par\parindent=16pt}

%
\def\numrefbk#1#2#3#4
   {\parindent=40pt
    \hang
    \noindent
    \rm {\hbox to 30truept{\hss #1\quad}}#2
   {\frenchspacing\sl #3\/}
    #4\par\parindent=16pt}

%

\def\ref#1{\par\noindent \hbox to 21pt{\hss
#1\quad}\frenchspacing\ignorespaces}

%
\def\frac#1#2{{#1 \over #2}}

%

%
\def\d{\hbox{\rm d}}

%
\def\e{\operatorname{e}}


\def\i{\operatorname{i}}
\chardef\ii="10

%

%

%

\catcode`\@=11
\def\vfootnote#1{\insert\footins\bgroup
    \interlinepenalty=\interfootnotelinepenalty
    \splittopskip=\ht\strutbox 
    \splitmaxdepth=\dp\strutbox \floatingpenalty=20000
    \leftskip=0pt \rightskip=0pt \spaceskip=0pt \xspaceskip=0pt
    \noindent\eightpoint\rm #1\ \ignorespaces\footstrut\futurelet\next\fo@t}

%
%
\def\eq(#1){\hfill\llap{(#1)}}
\catcode`\@=12
%
%



%
%





%
%

%
%

%
%

%

%

%

%
\def\gap{\;\lower3pt\hbox{$\buildrel > \over \sim$}\;}
%
%
\def\lap{\;\lower3pt\hbox{$\buildrel < \over \sim$}\;}
\def\tqs{\hbox to 25pt{\hfil}}


%

{\obeylines\gdef\startdisplay#1
  {\catcode`\^^M=5$$#1\halign\bgroup\indent##\hfil&&\qquad##\hfil\cr}}
\outer\def\enddisplay{\crcr\egroup$$}

\chardef\other=12
\def\ttverbatim{\begingroup \catcode`\\=\other \catcode`\{=\other
  \catcode`\}=\other \catcode`\$=\other \catcode`\&=\other
  \catcode`\#=\other \catcode`\%=\other \catcode`\~=\other
  \catcode`\_=\other \catcode`\^=\other
  \obeyspaces \obeylines \tt}
{\obeyspaces\gdef {\ }}  

\outer\def\begintt{$$\let\par=\endgraf \ttverbatim \parskip=0pt
  \catcode`\|=0 \rightskip=-5pc \ttfinish}
{\catcode`\|=0 |catcode`|\=\other 
  |obeylines 
  |gdef|ttfinish#1^^M#2\endtt{#1|vbox{#2}|endgroup$$}}

\catcode`\|=\active
{\obeylines\gdef|{\ttverbatim\spaceskip=.5em plus.25em minus.15em
                                            \let^^M=\ \let|=\endgroup}}%


\TagsOnRight

\tracingstats=1    

\font\twelverm=cmr10 scaled 1200
\font\twelvebf=cmbx10 scaled 1200
\normalbaselineskip=15pt
\baselineskip=14pt

\def\CD{{\Cal D}}

\def\ih{{\i\over\hbar}}
\def\viert{{1\over4}}
\def\half{{1\over2}}
\def\bFj{ {\bar F}_{(j)} }
\def\bFjp{ \bFj' }
\def\bFjpd{ \bFj''' }
\def\bFjpz{ \bFj^{\prime\,2} }
\def\bFjpv{ \bFj^{\prime\,4} }
\def\bFjppz{ \bFj^{\prime\prime\,2}  }
\def\bFjd{ {\dot{\bar F}_{(j)}} }
\def\bgj{ {\bar g}_{(j)} }
\def\bgjp{ \bgj' }
\def\bgjpz{ \bgj^{\prime\,2} }
\def\bgjd{ {\dot{\bar g}_{(j)}} }
\def\lTime{{t'}}
\def\uTime{{t''}}
\def\artan{\operatorname{arctan}}
\def\artanh{\operatorname{artanh}}
\def\erfc{\operatorname{erfc}}
\def\myalign{\allowdisplaybreaks\align}

\newcount\glno
\def\plus{\advance\glno by 1}
\def\num{\the\glno}

\newcount\refno
\def\add{\advance\refno by 1}
\refno=1

\edef\FH{\the\refno}\add
\edef\DURU{\the\refno}\add
\edef\GROGO{\the\refno}\add
\edef\GROS{\the\refno}\add
\edef\GOOa{\the\refno}\add
\edef\DMN{\the\refno}\add
\edef\ROSP{\the\refno}\add
\edef\MADE{\the\refno}\add
\edef\DEPO{\the\refno}\add
\edef\DALCH{\the\refno}\add
\edef\BK{\the\refno}\add
\edef\LR{\the\refno}\add
\edef\DK{\the\refno}\add
\edef\PAKS{\the\refno}\add
\edef\STEI{\the\refno}\add
\edef\GRSb{\the\refno}\add
\edef\KLEI{\the\refno}\add
\edef\STORCH{\the\refno}\add
\edef\PELST{\the\refno}\add
\edef\YODEWM{\the\refno}\add
\edef\CAST{\the\refno}\add
\edef\FLM{\the\refno}\add
\edef\PI{\the\refno}\add
\edef\GRSc{\the\refno}\add
\edef\BLIND{\the\refno}\add
\edef\PAU{\the\refno}\add


{\nopagenumbers
\pageno=0
\centerline{January 1993\hfill SISSA/2/93/FM}
\vskip1cm
\centerline{\twelvebf PATH INTEGRAL SOLUTION OF A CLASS OF EXPLICITLY}
\medskip
\centerline{\twelvebf TIME-DEPENDENT POTENTIALS}
\bigskip
\centerline{\twelverm CHRISTIAN GROSCHE}
\bigskip
\centerline{\it Scuola Internazionale Superiore di Studi Avanzati}
\centerline{\it International School for Advanced Studies}
\centerline{\it Via Beirut 4, 34014 Trieste, Miramare, Italy}
\vfill
\midinsert
\narrower
\noindent
{\bf Abstract.}
A specific class of explicitly time-dependent potentials is studied by
means of path integrals. For this purpose a general formalism to treat
explicitly time-dependent space-time transformations in path integrals
is sketched. An explicit time-dependent model under consideration is of
the form $V(q,t)=V[q/\zeta(t)]/\zeta^2(t)$, where $V$ is a usual
potential, and $\zeta(t)=(at^2+2bt+c)^{1/2}$. A recent result of Dodonov
 et al.\ for calculating corresponding propagators is incorporated into
the path integral formalism by performing a space-time transformation.
Some examples illustrate the formalism.
\endinsert
\bigskip
\centerline{\vrule height0.25pt depth0.25pt width4cm\hfill}
\noindent
{\sevenrm $^*$ Address from August 1993: II.Institut f\"ur Theoretische
          Physik, Universit\"at Hamburg, Luruper Chaussee 149,
          22761 Hamburg, Germany.}
\eject}


\glno=0                      
\line{\bf 1.\ Introduction\hfill}
\medskip\noindent
Explicitly time-dependent problems are of great importance in quantum
mechanics, in particular in scattering theory, in cosmology, in systems
with a time varying force field, or for the investigation of excitation
spectra.  However, there are only a few available exact solutions and
even less discussions in the context of path integrals; here e.g.\ the
famous forced harmonic oscillator [\FH] must be mentioned, reference
[\DURU] for a specific class of explicitly time-dependent
one-dimensional problems, and the general
quadratic Lagrangian with time-dependent coefficients [\GROGO, \GROS],
as well as the time-dependent radial harmonic oscillator [\GOOa]. As a
matter of fact, general formul\ae\ are difficult to achieve and usually
only special cases seem to allow to state explicit solutions, e.g.\ the
``moving potentials'' of reference [\DURU].

Recently, Dodonov et al.\ [\DMN] (and shortly later on Rogers and
Spector [\ROSP]) have discussed a further class of explicitly
time-dependent potentials. Of particular interest is the model of a
particle moving inside an infinite square well of moving width, a model
important in cosmology, see e.g.\ Makowski and Dembinski [\MADE],
Devoto and Pomori\v sac [\DEPO], and Da Luz and Cheng [\DALCH] and
references therein.

The general structure of all these problems is that the corresponding
quantum Hamiltonian has the following form
\plus$$H={p^2\over2m}+{1\over\zeta^2(t)}V\bigg({x\over\zeta(t)}\bigg)
  \enspace,
  \tag\num$$
\edef\numa{\num}%
with $x$ the spatial variable, $p$ its conjugate momentum, $\zeta(t)
=(at^2+2bt+c)^{1/2}$, and $a,b,c$ some constants. In reference [\DMN]
this kind of time-dependent potentials was solved by means of a
particular integral of motion [\BK, \LR]
\plus$$I(x,t)=\zeta^2(t)H-{\d\over\d t}{\zeta^2(t)\over4}(xp+px)
  +mx^2{\d^2\over\d t^2}{\zeta^2(t)\over4}\enspace,
  \tag\num$$
by looking for eigenfunctions of the operator $I$, i.e.\ $I(x,t)\Psi(x,
t)=E\Psi(x,t)$. Having found the eigenfunctions $\Psi(x,t)$ it is an
easy task to construct the corresponding propagator, required, the
propagator is known for the time-independent version of $H$ in
Eq.~(\numa), plus an additional harmonic potential.

{}From the point of view of path integrals, this is an indirect reasoning
and by no means satisfactory. In this Letter I want to show that the
corresponding kind of time-dependent problems according to Eq.~(\numa)
can be done by a space-time transformation.


\bigskip\noindent
\line{\bf 2.\ Explicitly Time-Dependent Duru--Kleinert
      Transformation\hfill}
\medskip\noindent
In order to discuss explicitly time-dependent space-time transformations
we start by considering the usual path integral formulation according to
\hfuzz=5pt
\plus$$\myalign
  K&(x'',x';\uTime,\lTime)
  =\int\limits_{x(\lTime)=x'}^{x(\uTime)=x''}\CD x(t)
   \exp\left\{\ih\int_\lTime^\uTime\bigg[{m\over2}\dot x^2
     -V(x)\bigg]dt\right\}
  \\   &
  =\lim_{N\to\infty}\bigg({m\over2\pi\i\epsilon\hbar}\bigg)^{N/2}
    \prod_{j=1}^{N-1}\int dx_j
    \exp\left\{\ih\sum_{j=1}^N\bigg[{m\over2\epsilon}(\Delta x_j)^2
    -\epsilon V(x_j)\bigg]\right\}\enspace.
  \tag\num\endalign$$
\edef\numi{\num}%
\hfuzz=3pt
Here $\Delta x_j=x_j-x_{j-1}$, $x_j=x(t_j)$,
$t_j=t'+j\epsilon$, $\epsilon=T/N$, $T=t''-t'$ fixed, and I have
used standard notation for path integrals [\FH]. In order to avoid
cumbersome notation I only consider the one-dimensional case. I
consider an explicitly time-dependent coordinate transformation
according to $x=h(q,t)$. Implementing this transformation one has to
keep all terms of $O(\epsilon)$ in the lattice definition of the path
integral (\numi), and expands about midpoints $\bar q_j=
\half(q_j+q_{j-1})$, $\bar t_j=\half(t_j+t_{j-1})$. The
measure is transformed according to
\plus$$\myalign
    \prod_{j=1}^{N-1}dx_j
  =&\prod_{j=1}^{N-1}h'(q_j,t_j)dq_j
  \\
  =&[h'(q'',\uTime)h'(q',\lTime)]^{-1/2}
  \prod_{j=1}^{N}[h'(q_{j-1},t_{j-1})h'(q_j,t_j)]^{1/2}
  \prod_{j=1}^{N-1}dq_j
  \\
  \dot=&[h'(q'',\uTime)h'(q',\lTime)]^{-1/2}
  \prod_{j=1}^N \bFjp
  \left[1+{\i\epsilon\hbar\over8m\bFjpz}
  \left({\bFjpd\over\bFjp}-{\bFjppz\over\bFjpz}\right)\right]
  \prod_{j=1}^{N-1}dq_j\enspace.
  \tag\num\endalign$$
Here denote $\bFj=h(\bar q_j,\bar t_j)$ etc., and
$h'(q,t)=\partial h(q,t)/\partial q$,
$\dot h(q,t)=\partial h(q,t)/\partial t$.
For the kinetic term in the exponential we get
\plus$$\multline
   \exp\bigg[{\i m\over2\epsilon\hbar}(x_j-x_{j-1})^2\bigg]
  \\   \dot=
  \exp\Bigg\{{\i m\over2\hbar}
   \Bigg[{\bFjpz\over\epsilon}(q_j-q_{j-1})^2
   +\epsilon{\bFjd}^2+2\bFjp\bFjd\Delta q_j\Bigg]
   -{\i\epsilon\hbar\over8m\bFjpz}{\bFjpd\over\bFjp}\Bigg\}\enspace.
  \endmultline
  \tag\num$$
This gives the coordinate transformation formula
$$\myalign
  K&\big(h(q'',\uTime),h(q',\lTime);\uTime,\lTime\big)
  \\   &
  =[h'(q'',\uTime)h'(q',\lTime)]^{-1/2}
  \lim_{N\to\infty}\bigg({m\over2\pi\i\epsilon\hbar}\bigg)^{N/2}
  \prod_{j=1}^{N-1}\int dq_j\cdot\prod_{j=1}^N \bFjp
  \\   &\qquad\times
  \exp\Bigg\{\ih\sum_{j=1}^N\Bigg[{m\over2\epsilon}
  \bFjpz(\Delta q_j)^2
    +{m\over2}\Big(\epsilon{\bFjd}^2+2\bFjp\bFjd\Delta q_j\Big)
  \\   &\qquad\qquad\qquad\qquad\qquad\qquad\qquad\qquad
  -\epsilon V(\bFj)
  -{\epsilon\hbar^2\over8m}{\bFjppz\over\bFjpv}\Bigg]\Bigg\}
  \tag\num\\   \global\plus   &
  \equiv[h'(q'',\uTime)h'(q',\lTime)]^{-1/2}
  \int\limits_{q(\lTime)=q'}^{q(\uTime)=q''}h'(q,t)\CD_{MP}q(t)
  \\   &\qquad\times
   \exp\Bigg\{\ih\int_\lTime^\uTime
   \Bigg[{m\over2}\bigg({h'}^2(q,t)\dot q^2
   +{\dot h}^2(q,t)+2h'(q,t)\dot h(q,t) \dot q\bigg)
  \\   &\qquad\qquad\qquad\qquad\qquad\qquad\qquad\qquad
  -V(h(q,t))-{\hbar^2\over8m}{{h''}^2(q,t)\over{h'}^4(q,t)}\bigg)
  \Bigg]dt\Bigg\}\enspace.
  \tag\num\endalign$$
\edef\numj{\num}%
The notation $\int\CD_{MP}q$ means that the path integral is
{\it defined\/} on mid-points. It is obvious that the path integral
(\numj) is short of being completely satisfactory. Whereas the
transformed potential $V(h(q,t))$ may have a convenient form when
expressed in the new coordinate $q$, the kinetic term ${m\over2}
{h'}^2\dot q^2$ is in general nasty, and the term linear in $\dot q$
 also cannot be seen as of practical use. To achieve a more convenient
form of the path integral (\numj) I proceed in two steps, first, the
term proportional to $\dot q$ is gauged away, second, a
time-transformation is performed ([\DK-\PELST], in particular
Refs.~[\STORCH, \PELST], and c.f.\ [\PELST] for a comprehensive but
slightly different discussion)

To deal with the $\dot q$ term we introduce the identity and expand
it about midpoints
\plus$$\myalign
  1&={g(q'',\uTime)\over g(q',\lTime)}
  \prod_{j=1}^N{g(q_{j-1},t_{j-1})\over g(q_j,t_j)}
  \\   &
  \simeq{g(q'',\uTime)\over g(q',\lTime)}\prod_{j=1}^N
  \left(1-{\bgjp\over\bgj}\Delta q_j
  +{\Delta^2q_j\over2}{\bgjpz\over{\bgj}^2}
  -\epsilon{\bgjd\over\bgj}\right)
  \\   &
  \simeq{g(q'',\uTime)\over g(q',\lTime)}
  \exp\Bigg[-\sum_{j=1}^N\Bigg({\bgjp\over\bgj}\Delta q_j
   +\epsilon{\bgjd\over\bgj}\Bigg)\Bigg]\enspace,
  \tag\num\endalign$$
\edef\numo{\num}%
due to $\e^{-z}\simeq1-z+z^2/2$, ($\vert z\vert\ll1$), and hence no term
$\propto(\Delta q_j)^2$ is present. Here $g(q,t)$ denotes a
to-be-determined function in $q$ and $t$; by the symbol ``$\simeq$'' we
have denoted that we are keeping terms up to $O(\epsilon)\propto
O\big((\Delta q_j)^2\big)$. Implementing the identity (\numo) into the
path integral (\numj) yields
\plus$$\myalign
  K&\big(h(q'',\uTime),h(q',\lTime);\uTime,\lTime\big)
  \\   &
  =[h'(q'',\uTime)h'(q',\lTime)]^{-1/2}{g(q'',\uTime)\over g(q',\lTime)}
  \int\limits_{q(\lTime)=q'}^{q(\uTime)=q''}h'(q,t)\CD_{MP}q(t)
  \\   &\qquad\times
   \exp\Bigg\{\ih\int_\lTime^\uTime
   \Bigg[{m\over2}\Big({h'}^2(q,t)\dot q^2+{\dot h}^2(q,t)\Big)
  +\bigg(mh'(q,t)\dot h(q,t)-{\hbar\over\i}{g'(q,t)\over g(q,t)}\bigg)
  \dot q
  \\   &\qquad\qquad\qquad\qquad\qquad
  -{\hbar\over\i}{\dot g(q,t)\over g(q,t)}
  -V(h(q,t))-{\hbar^2\over8m}{{h''}^2(q,t)\over{h'}^4(q,t)}
  \Bigg]dt\Bigg\}\enspace.
  \tag\num\endalign$$
\edef\numm{\num}%
To gauge away the $\dot q$-term we choose the function $g(q,t)$ in
such a way that
\plus$${\i m\over\hbar}h'(q,t)\dot h(q,t)={g'(q,t)\over g(q,t)}\enspace,
  \tag\num$$
which gives for $g(q,t)$ the solution
\plus$$g(q,t)=\exp\Bigg({\i m\over\hbar}\int^q \!\!h'(z,t)\dot
  h(z,t)dz\Bigg)
  \enspace.
  \tag\num$$
\edef\numq{\num}%
Insertion into the path integral (\numm) then gives
\plus$$K\big(h(q'',\uTime),h(q',\lTime);\uTime,\lTime\big)
  =A(q'',q';\uTime,\lTime)
  \widetilde K\big(h(q'',\uTime),h(q',\lTime);\uTime,\lTime\big)
  \enspace,
  \tag\num$$
with the prefactor $A(\uTime,\lTime)$
\plus$$A(q'',q';\uTime,\lTime)
  =\exp\Bigg[{\i m\over\hbar}\Bigg(
      \int^{q''}\! h'(z,\uTime)\dot h(z,\uTime)dz
     -\int^{q'}\! h'(z,\lTime)\dot h(z,\lTime)dz\Bigg)\Bigg]\enspace,
  \tag\num$$
and the path integral $\widetilde K(\uTime,\lTime)$ is given by
\plus$$\multline
  \widetilde K\big(h(q'',\uTime),h(q',\lTime);\uTime,\lTime\big)
  =[h'(q'',\uTime)h'(q',\lTime)]^{-1/2}
  \\   \qquad\times
   \int\limits_{q(\lTime)=q'}^{q(\uTime)=q''}h'(q,t)\CD_{MP}q(t)
  \exp\Bigg\{\ih\int_\lTime^\uTime\Bigg[
  {m\over2}\Big({h'}^2(q,t)\dot q^2+\dot h^2(q,t)\Big)-V(h(q,t))
  \hfill\\
  -{\hbar^2\over8m}{{h''}^2(q,t)\over{h'}^4(q,t)}
  -m\int^q\!\!\Big(h'(z,t)\ddot h(z,t)+\dot h(z,t)\dot h'(z,t)\Big)dz
   \Bigg]dt\Bigg\}\enspace.
  \endmultline
  \tag\num$$
\edef\numk{\num}%
For the case $\dot h'(q,t)\not=0$ this can simplified into
\plus$$\multline
  \widetilde K\big(h(q'',\uTime),h(q',\lTime);\uTime,\lTime\big)
  =[h'(q'',\uTime)h'(q',\lTime)]^{-1/2}
  \\   \qquad\times
   \int\limits_{q(\lTime)=q'}^{q(\uTime)=q''}h'(q,t)\CD_{MP}q(t)
  \exp\Bigg\{\ih\int_\lTime^\uTime\Bigg[{m\over2}{h'}^2(q,t)\dot q^2
  -V(h(q,t))
  \hfill\\
  -{\hbar^2\over8m}{{h''}^2(q,t)\over{h'}^4(q,t)}
  -m\int^q\!\! h'(z,t)\ddot h(z,t)dz\Bigg]dt\Bigg\}\enspace.
  \endmultline
  \tag\num$$
\edef\numl{\num}%
It is also obvious that Eqs.~(\numk,\numl) can be generalized to
the $D$ dimensional case.

Proceeding, one first for the resolvent makes use of the identity
\plus$${1\over\hat H-\hat E}=\hat f_r(q,t)
  {1\over\hat f_l(q,t)(\hat H-\hat E)\hat f_r(q,t)}\hat f_l(q,t)
  \enspace,
  \tag\num$$
where $\hat H$ is the Hamiltonian corresponding to the path integral
$K(\uTime,\lTime)$, $\hat f_{l,r}(q,t)$ are multiplication operators
in $q$ and $t$, multiplying from the left, respectively from the right,
onto the operator $(\hat H-\hat E)$, and and $\hat E=\i\hbar\partial_t$
is the energy operator, and second introduces
a new pseudo-time $s''$ defined by [\DK-\PELST]:
\plus$$s''\equiv\tau(\uTime)=\int_\lTime^\uTime
  {dt\over f_l(q,t)f_r(q,t)}\enspace.
  \tag\num$$
\edef\numn{\num}%
Introducing the matrix element [\KLEI, \PELST] $<t\vert E>=\e^{-\i
Et/\hbar}/\sqrt{2\pi\hbar}$ with the corresponding representation
of $\delta(t''-t'-s)=<t''\vert\e^{\i\hat E s/\hbar}\vert t'>$, together
with the completeness relation $\int dp\int dE\vert p,E><p,E\vert=1$,
we obtain for the path integral $K(t'',t')$
\plus$$\myalign
  K\big(&h(q'',\uTime),h(q',\lTime);\uTime,\lTime\big)
  \\   &
  =f_r(x'',\uTime)f_l(x',\lTime)
  \Big[h'(q'',\uTime)h'(q',\lTime)\Big]^{-1/2}\int_0^\infty ds''
  \\   &\qquad\times
  \lim_{N\to\infty}\prod_{j=1}^{N-1}\int dq_j\int dt_j
  \left[ \prod_{j=1}^N\delta(\Delta t_j-\epsilon_s f_{l,j}f_{r,j-1})
  \sqrt{ {m\over2\pi\i\epsilon_s\hbar}\cdot
         {h'_jh'_{j-1}\over f_{l,j}f_{r,j-1}} }\,\right]
  \\   &\qquad\times
  \exp\left\{\ih\sum_{j=1}^N\bigg[{m\over2\epsilon_s}
  \cdot{(h_j-h_{j-1})^2\over f_{l,j}f_{r,j-1}}
  -\epsilon_s f_{l,j}f_{r,j-1}V(h_j)\bigg]\right\}\enspace.
  \tag\num\endalign$$
\edef\nump{\num}%
Here denote $h_j=h(q_j,t_j)$, etc., and $\epsilon_s$ and
$\epsilon$ are related by $\epsilon=\epsilon_s f_{l,j}f_{r,j-1}$.
We make the choice $f_{l,j}=h'_j$, $f_{r,j-1}=h'_{j-1}$, to
guarantee a symmetric transformation with respect to initial and final
coordinates, and again we expand about the midpoints $\bar q_j$ and
$\bar t_j$. We obtain similarly as before [identify $h(q,t(s))
\equiv h(q,s)$; $\dot h'(q,t)\not=0$, c.f.\ the remark following
Eq.~(\numk)].
\plus$$\myalign
  K\big(&h(q'',\uTime),h(q',\lTime);\uTime,\lTime\big)
  \\   &
  =\Big[h'(q'',\uTime)h'(q',\lTime)\Big]^{1/2}A(q'',q';\uTime,\lTime)
  \int_0^\infty ds''
  \\   &\qquad\times
  \lim_{N\to\infty}\prod_{j=1}^{N-1}\int dq_j\int dt_j
  \bigg({m\over2\pi\i\epsilon_s\hbar}\bigg)^{N/2}
  \prod_{j=1}^N\delta(\Delta t_j-\epsilon_s f_{l,j}f_{r,j-1})
  \\   &\qquad\times
  \exp\left\{\ih\sum_{j=1}^N\bigg[{m\over2\epsilon_s}(\Delta q_j)^2
  -\epsilon_s \bFjpz V(\bFj)
  -\epsilon_s\Delta V(\bar q_j,\bar s_j)\bigg]\right\}\enspace.
  \tag\num\endalign$$
Here $\Delta V$ denotes the quantum potential
\plus$$\Delta V(q,s)={\hbar^2\over8m}\bigg(
   3{{h''}^2(q,s)\over{h'}^2(q,s)}-2{h'''(q,s)\over h'(q,s)}\bigg)
  +m{h'}^2(q,s)\int^q\!\! h'(z,s)\ddot h(z,s)dz\enspace.
  \tag\num$$
The $\delta$-function integrations successively determine the
values of $t_j$ by means of an iteration process
\plus$$t_j-t_{j-1}=\epsilon_sh'_jh'_{j-1}
            =\epsilon_s{h'}^2_j+O(\epsilon^2)\enspace,
  \tag\num$$
which allows the actual iterated integration and evaluation of the
argument of the $\delta$-function, respectively, where the
$O(\epsilon_s^2)$ can be ignored, yielding [\PELST]
\plus$$t-t_j=\sum_{k=j}^N\epsilon_s{h'}^2_k\enspace.
  \tag\num$$
Because there is one more $\delta$-function as integration, the last
one is expanded into its Fourier representation and thus we arrive
finally for the combined transformations $x=h(q,T)$ and $dt/f_r(q,t)
f_l(q,t)=dt/{h'}^2(q,t)=ds$ at the space-time (Duru-Kleinert)
transformation formul\ae
$$\myalign
  K(x'',x';\uTime,\lTime)
  &=\Big[h'(q'',\uTime)h'(q',\lTime)\Big]^{1/2}
    A(q'',q';\uTime,\lTime)
  \\    &\qquad\qquad\qquad\qquad\times
    \int_{-\infty}^\infty{dE\over2\pi\i}\e^{-\i ET/\hbar}
    G(q'',q';E)
  \tag\num\\   \global\plus
  G(q'',q';E)
  &=\ih\int_0^\infty \widehat K(q'',q';s'')ds''
  \tag\num\endalign$$
with the path integral $\widehat K(s'')$ given by
\plus$$\multline
  \!\!\!\!
  \widehat K(q'',q',s'')
  \\   \hfill
  =\int\limits_{q(0)=q'}^{q(s'')=q''}\CD_{MP}q(s)
  \exp\Bigg\{\ih\int_0^{s''}\Bigg[{m\over2}\dot q^2
  -{h'}^2(q,s)\Big(V(h(q,s))-E\Big)-\Delta V(q,s)
  \Bigg]ds\Bigg\}\enspace.
  \\ \ \endmultline
  \tag\num$$
Note that in comparison to Refs.~[\STORCH, \PELST] there is no potential
term $\propto-\i\hbar h'(q,t)\dot h'(q,t)$ present, c.f.\ the model in
the next Section and the discussion in the summary.
It must be noted that the whole procedure as sketched here remains
only on a formal level, however with well-defined rules.
Attempts to put it on a more sound mathematical basis can be
e.g.\ found Ref.~[\YODEWM], and more recently c.f.\ Refs.~[\CAST, \FLM].


\bigskip\noindent
\line{\bf 3.\ The Model\hfill}
\medskip\noindent
Let us consider the path integral formulation corresponding to
Eq.~(\numa)
\plus$$K(x'',x';\uTime,\lTime)
  =\int\limits_{x(\lTime)=x'}^{x(\uTime)=x''}\CD x(t)
   \exp\left\{\ih\int_\lTime^\uTime\left[{m\over2}\dot x^2
     -{1\over\zeta^2(t)}V\bigg({x\over\zeta(t)}\bigg)\right]dt\right\}
  \enspace.
  \tag\num$$
\edef\numb{\num}%
In order to discuss the path integral (\numb), I consider in the first
step the (time-dependent) coordinate transformation $x(t)=\zeta(t)
q(t)$ according to the rules in Section~2. Inserting this transformation
into $\Delta V$ and the prefactor $A(t'',t')$ yields (note that we also
can perform a partial integration in the Lagrangian in the exponential
without the delay of $\Delta V$, because the transformation is linear in
the spatial coordinates, $D$ the spatial dimension, $\zeta'=\zeta(t')$,
$\zeta''=\zeta(t'')$, etc.)
\plus$$K(x'',x';\uTime,\lTime)
  =\big(\zeta''\zeta'\big)^{-D/2}
  \exp\left[{\i m\over2\hbar}\left({x''}^2{\dot\zeta''\over\zeta''}
                     -{x'}^2{\dot\zeta'\over\zeta'}\right)\right]
  \widetilde K (q'',q';\uTime,\lTime)\enspace,
  \tag\num$$
\edef\numd{\num}%
with the path integral $\widetilde K(\uTime)$ given by
\plus$$\myalign
               &{\widetilde K}(q'',q';\uTime,\lTime)
  =\lim_{N\to\infty}\bigg({m\zeta_j\zeta_{j-1}
                     \over2\pi\i\epsilon\hbar}\bigg)^{ND/2}
    \prod_{j=1}^{N-1}\int dq_j
  \\           &\qquad\times
    \exp\left\{\ih\sum_{j=1}^N\left[{m\over2\epsilon}
    \zeta_j\zeta_{j-1}(\Delta q_j)^2
    -{\epsilon\over\zeta_j\zeta_{j-1}}
     \bigg({m\over2}{\omega'}^2q_j^2
               +V(q_j)\bigg)\right]\right\}\enspace.
  \tag\num\endalign$$
with ${\omega'}^2=ac-b^2$. The path integral (\numd) together to
$\widetilde K(\uTime,\lTime)$ corresponds exactly to the path integral
(\numk).

Proceeding we perform a time-transformation $dt/\zeta^2(t)=ds$, i.e.\
we set $f^2(s)=\zeta^2[t(s)]$, with a new time $s''$ defined by
\plus$$s''\equiv\tau(\uTime)=\int_\lTime^\uTime{dt\over\zeta^2(t)}
  \left\{\alignedat 3
  &={1\over\omega'}\artan{at+b\over\omega'}\bigg\vert_\lTime^\uTime
  &\qquad  &({\omega'}^2>0)\enspace,
  \\
  &=-{1\over\vert\omega'\vert}\artanh{at+b\over\vert\omega'\vert}
  \bigg\vert_\lTime^\uTime
  &\qquad  &({\omega'}^2<0)\enspace,
  \\
  &={a\over b}{t\over at+b}\bigg\vert_\lTime^\uTime
  &\qquad  &({\omega'}^2=0)\enspace.
  \endalignedat\qquad\qquad\right\}
  \tag\num$$
This gives now the transformation formul\ae
$$\myalign
  \widetilde K(q'',q';\uTime,\lTime)
  &=\zeta(t')\zeta(t'')
    \int_{-\infty}^\infty {dE\over2\pi\i}G(q'',q';E)
   \e^{-\i ET/\hbar}
  \tag\num\\   \global\plus
  G(q'',q';E)&=\ih\int_0^\infty \widehat K(q'',q';s'')
  \exp\Bigg({\i E\over\hbar}\int_0^{s''}\zeta^2(t(s))ds\Bigg)ds''
  \enspace,
  \tag\num\endalign$$
\edef\numh{\num}%
and the path integral $\widehat K(s'')$ given by
\plus$$\widehat K(q'',q';s'')
  = \int\limits_{q(0)=q'}^{q(s'')=q''}\CD q(s)
   \exp\left\{\ih\int_0^{s''}\bigg[{m\over2}\dot q^2
   -{m\over2}{\omega'}^2q^2-V(q)\bigg]ds\right\}\enspace.
  \tag\num$$
\edef\numc{\num}%
This path integral has no explicit time-dependence, however an
additional harmonic part with frequency ${\omega'}^2=ac-b^2$ is present.
Let us assume that we can write down the solution of the path integral
(\numc) and call it $K_{\omega',V}(s'')$. We obtain [c.f.\
Eqs.~(\numn,\nump)]
\plus$$\align
  \widetilde K&(x'',x';\uTime,\lTime)
  \\   &
  =\zeta(t')\zeta(t'')\int_{-\infty}^\infty{dE\over2\pi\hbar}
   \int_0^{s''}ds''\exp\bigg[{\i E\over\hbar}\bigg(\int_0^{s''}
   \zeta^2(t(s))ds-T\bigg)\bigg] K_{\omega',V}(q'',q';s'')
  \\   &
  =K_{\omega',V}
  \bigg({x''\over\zeta''},{x'\over\zeta'};\tau(\uTime)\bigg)
  \enspace,
  \tag\num\endalign$$
and we therefore find for the propagator (\numd)
\plus$$K(x'',x';\uTime,\lTime)
 =\big(\zeta''\zeta'\big)^{-D/2}
  \exp\left[{\i m\over2\hbar}\left({x''}^2{\dot\zeta''\over\zeta''}
                     -{x'}^2{\dot\zeta'\over\zeta'}\right)\right]
  K_{\omega',V}
  \bigg({x''\over\zeta''},{x'\over\zeta'};\tau(\uTime)\bigg)
  \enspace,
  \tag\num$$
\edef\nume{\num}%
which is the result of reference [\DMN].

Let us note that for time-dependent potential problems according to
$V(x)\mapsto V(x-f(t))$ one derives from Eq.~(\numk) the identity
(c.f.\ [\DURU], $q'=x'-f'$, $f'=f(t')$, etc., note
$\dot f'(q,t)=0$)
\plus$$\multline
  \int\limits_{x(t')=x'}^{x(t'')=x''}\CD x(t)
  \exp\left\{\ih\int_{t'}^{t''}\bigg[
  {m\over2}\dot x^2-V(x-f(t))\bigg]dt\right\}
  \\   \qquad
  =\exp\left\{{\i m\over\hbar}\left[\dot f''(x''-f'')-\dot f'(x'-f')
  +\half\int_{t'}^{t''}\dot f^2(t)dt\right]\right\}
  \hfill\\   \qquad\qquad\times
  \int\limits_{q(t')=q'}^{q(t'')=q''}\CD q(t)
  \exp\left\{\ih\int_{t'}^{t''}\bigg[
  {m\over2}\dot q^2-V(q)-m\ddot f(t) q\bigg]dt\right\}\enspace.
  \hfill\endmultline
  \tag\num$$


\bigskip\noindent
\line{\bf 4.\ Examples\hfill}
\medskip\noindent
Let us discuss this result shortly for the two cases ${\omega'}^2>0$ and
${\omega'}^2<0$, respectively.
\newline
1) ${\omega'}^2\not=0$. For exactly solvable solutions of equation
(\nume) only harmonic potentials are relevant. Let us consider the
harmonic potential $V(x)=(m/2)\omega^2x^2$, and set $\Omega^2=
\omega^2+{\omega'}^2$ which may be positive or negative (for
$\Omega^2<0$ we have a harmonic repeller). Let $\Omega^2>0$. Then we
have the path integral identity ($D=1$)
\plus$$\myalign
  \int\limits_{x(\lTime)=x'}^{x(\uTime)=x''}&\CD x(t)
  \exp\left[{\i m\over2\hbar}\int_\lTime^\uTime\bigg(
     \dot x^2-{\omega^2\over\zeta^4(t)}x^2\bigg)dt\right]
  \\
  =&\big(\zeta'\zeta''\big)^{-1/2}
  \exp\left[{\i m\over2\hbar}\left({x''}^2{\dot\zeta''\over\zeta''}
                     -{x'}^2{\dot\zeta'\over\zeta'}\right)\right]
  \bigg({m\Omega\over2\pi\i\hbar\sin\Omega\tau(\uTime)}\bigg)^{1/2}
         \\   &\qquad\qquad\times
  \exp\left\{-{m\Omega\over2\i\hbar}\left[\left(
    {{x''}^2\over{\zeta''}^2}+{{x'}^2\over{\zeta'}^2}\right)
    \cot\Omega\tau(\uTime)
    -{2x'x''\over\zeta'\zeta''\sin\Omega\tau(\uTime)}\right]\right\}
  \enspace.\qquad
  \tag\num\endalign$$
\edef\numf{\num}%
Similarly we obtain for a radial harmonic potential
\plus$$V(r)={m\over2}\omega^2r^2+{\lambda^2-\viert\over2mr^2}
  \tag\num$$
the path integral identity $(\Omega$ and $\tau$ as before)
\plus$$\myalign
  \int\limits_{r(\lTime)=r'}^{r(\uTime)=r''}&\CD r(t)
  \exp\left[\ih\int_\lTime^\uTime\Bigg(
     {m\over2}\dot r^2-{m\over2}{\omega^2\over\zeta^4(t)}r^2
             -{\lambda^2-\viert\over2mr^2}\Bigg)dt\right]
         \\   &
  =\bigg({r'r''\over\zeta'\zeta''}\bigg)^{1/2}
  \exp\left[{\i m\over2\hbar}\left({r''}^2{\dot\zeta''\over\zeta''}
                     -{r'}^2{\dot\zeta'\over\zeta'}\right)\right]
  \bigg({m\Omega\over\i\hbar\sin\Omega\tau(\uTime)}\bigg)^{1/2}
         \\   &\qquad\times
  \exp\left[-{m\Omega\over2\i\hbar}\left(
    {{r''}^2\over{\zeta''}^2}+{{r'}^2\over{\zeta'}^2}\right)
    \cot\Omega\tau(\uTime)\right]
  I_\lambda\bigg({m\Omega r'r''\over\i\hbar\zeta'\zeta''
                 \sin\Omega\tau(\uTime)}\bigg)\enspace.
  \tag\num\endalign$$
\edef\numg{\num}%
Equations (\numf,\numg) can also be achieved by considering the explicit
 time-dependence $\Omega^2(t)=(\omega^2+{\omega'}^2) \zeta^{-4}(t)$ and
using the known solutions for the (radial) time-dependent harmonic
oscillator [\FH, \GOOa--\DMN, \PI], respectively.

\medskip\noindent
2) ${\omega'}^2=0$. Here we have reduced the number of parameters by
one and $\zeta(t)=(at+b)/\sqrt{a}\,$. For this special case we can
always rescale $\tau$ according to $\tau(\uTime)=\alpha T/a(1+\alpha
T)$ ($\alpha=b/a$) [\DMN, \ROSP], and every exactly solvable path
integral solution can be substituted into Eq.~(\numh), provided
$K_{0,V}\equiv K_V$ is known (see reference [\DMN] for examples, only
this particular case was treated in reference [\ROSP]).

We consider the example of the infinite well (IW) with one boundary
fixed at $x=0$, and the other moving uniformly in time according to
$L(t)=L_0 \zeta(t)$ [\DMN, \MADE-\BK]. The result then has the form
($\Theta(z,\tau)$ denotes a Jacobi-theta function)
\plus$$\multline
  K^{(IW)}(x'',x';\uTime,\lTime)
  ={(\zeta'\zeta'')^{-1/2}\over2L_0}
  \exp\left[{\i m\over2\hbar}\left({x''}^2{\dot\zeta''\over\zeta''}
                     -{x'}^2{\dot\zeta'\over\zeta'}\right)\right]
  \\    \times
  \Bigg[\Theta_3\bigg({x''/\zeta''-x'/\zeta'\over2L_0}
          ,-{\pi\hbar\tau(\uTime)\over2mL_0^2}\bigg)
   -\Theta_3\bigg({x''/\zeta''+x'/\zeta'\over2L_0}
          ,-{\pi\hbar\tau(\uTime)\over2mL_0^2}\bigg)\Bigg]\enspace.
  \endmultline
  \tag\num$$

For the second example we consider a time-dependent $\delta$-function
perturbation according to $V(x)=-\gamma\delta(x)/\zeta(t)$. We obtain
the path integral identity
\plus
$$\myalign
  \int\limits_{x(\lTime)=x'}^{x(\uTime)=x''}&\CD x(t)
  \exp\left[\ih\int_\lTime^\uTime\bigg({m\over2}
     \dot x^2+{\gamma\over\zeta(t)}\delta(x)\bigg)dt\right]
  \\
  =&\big(\zeta'\zeta''\big)^{-1/2}
  \exp\left[{\i m\over2\hbar}\left({x''}^2{\dot\zeta''\over\zeta''}
                     -{x'}^2{\dot\zeta'\over\zeta'}\right)\right]
         \\   &\quad\times
  \left\{
  \bigg({m\over2\pi\i\hbar\tau(\uTime)}\bigg)^{1/2}
  \exp\left[{\i m\over2\hbar\tau(\uTime)}
  \left({x''\over\zeta''}-{x'\over\zeta'}\right)^2\right]
  \right.\\   &\qquad\qquad
  +{m\gamma\over2\hbar^2}
   \exp\left[-{m\gamma\over\hbar^2}
  \left({\vert x''\vert\over\zeta''}+{\vert x'\vert\over\zeta'}\right)
  +\ih\tau(\uTime){m\gamma^2\over2\hbar^2}\right]
         \\   &\left.\qquad\qquad\qquad\qquad\times
  \erfc\Bigg[\sqrt{m\over2\i\hbar\tau(\uTime)}
  \left({\vert x''\vert\over\zeta''}+{\vert x'\vert\over\zeta'}
   -\ih\gamma\tau(\uTime)\right)\Bigg]\right\}
  \enspace.\qquad
  \tag\num\endalign$$

However, for numerous examples only $G_V(E)$ instead of $K_V(\uTime)$
can be explicitly stated, i.e.\ the energy-dependent Green function is
available [\GRSc]; here the best known example is the Coulomb-Green
function (see however [\DMN, \BLIND]). Then, instead of Eq.~(\nume),
only Eq.~(\numh) can be stated in closed form.


\bigskip
\line{\bf 5.\ Summary and Discussion\hfill}
\medskip\noindent
In this Letter I have studied a particular model of explicitly
time-dependent quantum mechanical problems by path integrals. In order
to do this I have sketched how to treat general explicitly
time-dependent space-time (Duru-Kleinert) transformation
(``process-cum-time substitution'' [\YODEWM]) in path integrals. In
comparison to, say, Ref.~[\PELST] there is no potential term $\propto
-\i\hbar h'(q,t)\dot h'(q,t)$ in the effective Lagrangian present.
An analysis of the result of Ref.~[\PELST] shows that this extra term
is due to another gauge, i.e.\ another function $g(q,t)$ is chosen.
Actually one has an additional factor $[h'(q,t)]^{1/2}$ in $g(q,t)$,
suggested by the there-used {\it postpoint\/} expansion. Changing the
gauge by taking $[h'(q,t)]^{-1/2}$ instead leads to a cancellation of
terms $\propto-\i\hbar h'(q,t)\dot h'(q,t)$. This difference can be
interpreted in the following way: In Ref.~[\PELST] the gauge has been
chosen in such a way that a term $\propto-\i\hbar h'(q,t)\dot h'(q,t)$
appears in the effective Lagrangian, hence an imaginary potential is
present. The imaginary potential can be on the one hand understood as a
source, respectively a sink for particles, because the transformation
of a time-independent Hamiltonian to a time-dependent one, say, has the
consequence that the new Hamiltonian does not conserve the energy;
this is now exactly balanced by the imaginary potential in order
to guarantee energy conservation of the entire (time-independent, say)
system. On the other, this term can be interpreted as a ``path-dependent
measure'' (as in [\PELST]), respectively another gauge is chosen, as in
this Letter; the latter case has the advantage that from the beginning
on the effects of the explicitly time-dependent Duru-Kleinert
transformation are incorporated in the weight factors of the
wave-functions. Actually, in our model, a time-dependent system is
tranformed into a time-independent one, and the transformed integration
measure is transfomed just in the correct way that it ``guarantees
that the probability density remains normalised in $D$-dimensional
space'' [\BK].

Therefore, the {\it midpoint\/} expansion leads in a very natural way
to the gauge as chosen in (\numq) by putting additional contributions
into the integration measure (Jacobean). Whereas in [\PELST] $g(q,t)$
is chosen in such a way that the corresponding Schr\"odinger equation
in the new coordinate $q$ and the pseudotime $s''$ does not have a
first order partial derivation, the present formalism allows this by
the introduction of a non-trivial momentum operator $p_q=-\i\hbar
(\partial_q+\half\Gamma_q)$ ($\Gamma_q=\partial_q\ln\sqrt{g}$, with
$\sqrt{g}\,dq$ the integration measure) [\GRSb, \PAU], which is
hermitean with respect to the inner product $\int\sqrt{g}\,
dqf^*(q)g(q)$. Furthermore, the postpoint expansion used in
Ref.~[\PELST] leads to an expansion into many terms of order $\epsilon$
which in the midpoint expansion are not present, respectively they are
cancelling each other. Therefore, our technique shows that the midpoint
prescription is far {\it simpler\/} in handling, in the {\it conceptual
understanding\/}, and gives {\it unambiguous results\/} in comparison
with already existing models [\DMN-\BK], in particular [\DMN, \BK].

Therefore we have shown that the technique of explicitly time-dependent
space-time transformation in path integrals provides the necessary tools
to treat explicitly time-dependent problems in a rigorous and explicit
way, leading to a general formula for the corresponding propagator. I
have unified the various approaches, and have presented a simpler
derivation of the time-dependent Duru--Kleinert transformation by means
of the midpoint prescription. I obtained a general formula for the
incorporation of this kind of explicit time dependence, provided the
propagator for the time-independent case is known. Whereas closed
expressions for $K_{\omega',V}$ are only possible for $V$ also
harmonic, and for $K_{0,V}$ there are only few, a formal spectral
expansion is always possible, i.e.\ yielding ($\omega'=0$, $D=1$)
\plus$$\myalign
  \int\limits_{x(\lTime)=x'}^{x(\uTime)=x''}&\CD x(t)
  \exp\left\{\ih\int_\lTime^\uTime\bigg[{m\over2}
  \dot x^2-{1\over\zeta^2(t)}V\bigg({x\over\zeta(t)}\bigg)
  \bigg]dt\right\}
         \\   &
  =\big(\zeta''\zeta'\big)^{-1/2}
  \exp\left[{\i m\over2\hbar}\left({x''}^2{\dot\zeta''\over\zeta''}
                     -{x'}^2{\dot\zeta'\over\zeta'}\right)\right]
         \\   &\qquad\times
  \int dE_\lambda\Psi_\lambda\bigg({x''\over\zeta''}\bigg)
                 \Psi_\lambda^*\bigg({x'\over\zeta'}\bigg)
  \exp\bigg(-{\i E_\lambda\over\hbar}
        \int_\lTime^\uTime{dt\over\zeta^2(t)}\bigg)\enspace,
  \tag\num\endalign$$
where $\int dE_\lambda$ denotes a Stieltjes integral to include bound
and scattering states $\Psi_\lambda$ with energy $E_\lambda$ of the
corresponding time-independent problem. This general result concludes
the discussion.

\newpage\noindent
\line{\bf Acknowledgement\hfill}
\medskip\noindent
I would like to thank the organizers of the graduate college,
``Quantenfeldtheorie und deren Anwendungen in der Elementarteilchen-
und Fest\-k\"orper\-physik'', University of Leipzig, Germany, where
part of this work was done, in particular B.\ Geyer, for their kind
hospitality. I also thank A.\ Pelster and A.\ Wunderlin for fruitful
discussions about explicitly time-dependent space-time transformations
in the path integral.


\bigskip\noindent
\line{\bf References\hfill}
\baselineskip=12pt
\medskip
\eightpoint
\eightrm
\item{[\FH]}
R.P.Feynman and A.Hibbs: Quantum Mechanics and Path Integrals
({\it McGraw Hill, New York}, 1965).
\item{[\DURU]}
I.H.Duru:
Quantum Mechanics of a Class of Time-Dependent Potentials;
{\it J.Phys.A: Math.Gen.}\ {\bf 22} (1989), 4827.
\item{[\GROGO]}
C.C.Grosjean and M.J.Goovaerts:
The Analytic Evaluation of One-Dimensional Gaussian Path Integrals;
{\it J.Comp.Appl.Math.}\ {\bf 21} (1988), 311.
\item{[\GROS]}
C.C.Grosjean:
A General Formula for the Calculation of Gaussian Path-Integrals in
Two and Three Euclidean Dimensions;
{\it J.Comput.Appl.Math.}\ {\bf 23} (1988), 199.
\item{[\GOOa]}
M.J.Goovaerts:
Path Integral Solution of Nonstationary Calogero Model;
{\it J.Math.Phys.}\ {\bf 16} (1975), 720.
\item{[\DMN]}
V.V.Dodonov, V.I.Man'ko and D.E.Nikonov:
Exact Propagators for Time-Dependent Coulomb, Delta and Other
Potentials;
{\it Phys.Lett.}\ {\bf A 162} (1992), 359.
\item{[\ROSP]}
J.Rogers and D.Spector:
Quantum Mechanics with Explicit Time Dependence;
{\it Phys.Lett.}\ {\bf A 170} (1992), 344.
\item{[\MADE]}
A.J.Makowski and S.T.Dembinski:
Exactly Solvable Models with Time-Dependent Boundary Conditions;
{\it Phys.Lett.}\ {\bf A 154} (1991), 217;
\newline
A.J.Makowski:
Two Classes of Exactly Solvable Quantum Models with Moving Boundary;
{\it J.Phys.A: Math.Gen.}\ {\bf A 25} (1992), 3419.
\item{[\DEPO]}
A.Devoto and B.Pomori\v sac:
Diffusion-Like Solutions of the Schr\"odinger Equation for a
Time-Dependent Potential Well;
{\it J.Phys.A: Math.Gen.}\ {\bf 25} (1992), 241.
\item{[\DALCH]}
M.G.E.Da Luz and B.K.Cheng:
Exact Propagators for Moving Hard-Wall Potentials;
{\it J.Phys.A: Math.Gen.}\ {\bf 25} (1992), L1043.
\item{[\BK]}
M.V.Berry and J.Klein:
Newtonian Trajectories and Quantum Waves in Expanding Force Fields;
{\it J.Phys.A: Math.Gen.}\ {\bf 17} (1984), 1805.
\item{[\LR]}
H.R.Lewis and W.B.Riesenfeld:
An Exact Quantum Theory of the Time-Dependent Harmonic Oscillator and of
a Charged Particle in a Time-Dependent Electromagnetic Field;
{\it J.Math.Phys.}\ {\bf 10} (1969), 1458.
\item{[\DK]}
I.H.Duru and H.Kleinert:
Solution of the Path Integral for the H-Atom;
{\it Phys.Lett.}\ {\bf B 84} (1979), 185;
Quantum Mechanics of H-Atoms From Path Integrals;
{\it Fort\-schr.Phys.}\ {\bf 30} (1982), 401.
\item{[\PAKS]}
N.K.Pak and I.S\"okmen:
General New-Time Formalism in the Path Integral;
{\it Phys.Rev.}\ {\bf A 30} (1984), 1629.
\item{[\STEI]}
F.Steiner:
Space-Time Transformations in Radial Path Integrals;
{\it Phys.Lett.}\ {\bf A 106} (1984), 356.
\item{[\GRSb]}
C.Grosche and F.Steiner:
Path Integrals on Curved Manifolds;
{\it Zeitschr.Phys.}\ {\bf C 36} (1987), 699.
\item{[\KLEI]}
H.Kleinert:
Path Integrals in Quantum Mechanics, Statistics and Polymer Phys\-ics
({\it World Scientific}, Singapore, 1990).
\item{[\STORCH]}
S.N.Storchak:
Rheonomic Homogeneous Point Transformation and Reparametrization in the
Path Integral;
{\it Phys.Lett.}\ {\bf 135} (1989) 77
\item{[\PELST]}
A.Pelster and A.Wunderlin:
On the Generalization of the Duru-Kleinert-Propaga\-tor Transformations;
{\it Zeitschr.Phys.}\ {\bf B 89} (1992), 373.
\item{[\YODEWM]}
A.Young and C.DeWitt-Morette,
Time Substitutions in Stochastic Processes as a Tool in Path
Integration;
{\it Ann.Phys.(N.Y.)} {\bf 169} (1986), 140.
\item{[\CAST]}
D.P.L.Castrigiano and F.St\"ark:
New Aspects of the Path Integrational Treatment of the Coulomb
Potential;
{\it J.Math.Phys.}\ {\bf 30} (1989), 2785.
\item{[\FLM]}
W.Fischer, H.Leschke and P.M\"uller:
Changing Dimension and Time: Two Well-Founded and Practical Techniques
for Path Integration in Quantum Physics;
{\it J.Phys.A: Math.Gen.}\ {\bf 25} (1992), 3835.
\item{[\PI]}
D.Peak and A.Inomata:
Summation Over Feynman Histories in Polar Coordinates;
{\it J.Math.Phys.}\ {\bf 10} (1969), 1422.
\item{[\GRSc]}
C.Grosche and F.Steiner:
Table of Feynman Path Integrals,
to appear in {\it Springer Tracts in Modern Physics}.
\item{[\BLIND]}
S.M.Blinder:
Analytic Form of the Nonrelativistic Coulomb Propagator;
{\it Phys.Rev.}\ {\bf A 43} (1991), 13.
\item{[\PAU]}
W.Pauli: Die allgemeinen Prinzipien der Wellenmechanik,
in S.Fl\"ugge: Handbuch der Physik, Band V/1
({\it Springer-Verlag}, Berlin, 1958)



\enddocument